\documentclass[12pt,showpacs,prc]{revtex4}
\usepackage{epsfig,amssymb,amsfonts}
\usepackage{graphicx}
\begin{document}

\draft

\title{Radial flow has little effect on clusterization at intermediate energies in the framework of the Lattice Gas Model }

\author
{
C. B. Das$^{1,2}$, L. Shi$^1$ and S. Das Gupta$^1$
}

\affiliation{$^1$Physics Department, McGill University, 
Montr{\'e}al, Canada H3A 2T8 \\
$^2$Physics Division, Variable Energy Cyclotron Center, 
1/AF Bidhannagar, Kolkata 700 064, India}

\date{\today}

\begin{abstract}
{
The Lattice Gas Model was extended to incorporate the effect of radial flow. Contrary to popular belief, radial flow has little effect on the clusterization process in intermediate energy heavy-ion collisions except adding an ordered motion to the particles in the fragmentation source. We compared the results from the lattice gas model with and without radial flow to experimental data. We found that charge yields from central collisions are not significantly affected by inclusion of any reasonable radial flow.
}
\end{abstract}

%
%
%

\maketitle
                      
%
%
%
%

Assuming local equilibrium and ignoring any collective motions, many statistical models are used to explain the cluster formation process in multifragmentations. 
Ignoring certain collective motions, such as sideward flow, is often justified by the short range nature of correlations which dominates the cluster formation process. Sideward flow only adds a uniform boost to the momentum of the particles in the neighborhood of a given particle and thus does not change the phase space distributions in the local rest frame. However, essential difference arises when the collective motion adds a nonuniform boost to the particles in the the neighborhood of a given particle. Such nonuniform boost could not be cancelled by a simple change of reference frame, and could induce a nontrivial change in the phase space distribution. One of the simple nonuniform boosts is radial flow \cite{Siemens}, and we will consider the effects of it on the cluster production process.

Experimentally, we know there is a strong radial flow in central heavy-ion collision. The radial flow could often account for half of the kinetic energy of the particles. In the paper \cite{Kunde1995}, essentially different pattern of charge yields are found in central collisions than in peripheral collisions. Kunde {\em et al}. \cite{Kunde1995} compared two systems with similar thermal energy (at least the authors of \cite{Kunde1995} claim so), but quite different radial flow energy. One of the systems is from low energy central collisions, where significant radial flow is present. Another system is from higher energy peripheral collisions where little radial flow is present. The charge yields from the central collisions are found to be of exponential type while those from the peripheral collisions are of power law type. Whether such a strong effect for radial flow exists or not has become a key issue for the understanding of cluster formations in intermediate energy heavy-ion collisions. 

For some simplified statistical models, such as Copenhagen model \cite{Bondorf:1995,Bondorf:1985a,Bondorf:1985b,Bondorf:1985c}, the Berlin 
model \cite{Gross:1997,Gross:1985,Gross:1987a,Gross:1987b} and the thermodynamic canonical model \cite{Dasgupta,Chase:1993nw,Chase:1995ad,Das:2003}, the radial flow can be added only a posteriori. Thus, the radial flow is not included at a fundamental level. Transport models, such as BUU and QMD could incorporate the flow features naturally, but have trouble to reproduce the details of charge yields \cite{Danielewicz:1991,Colonna:1998,Dorso:1993,Puri:1998te,Chernomoretz:2004}. In this sense, the lattice gas model bears some advantages in being able to incorporate flow in a natural way and also simulates some fine details of cluster productions.  

In a previous paper, we have developed the lattice gas model to include radial flow \cite{Das2001}. It was also applied to a small system with mass $A=84$ to test the effect of radial flow. The radial flow incorporated in the calculation is quite small, on the order of $2$ AMeV, and the effect of radial flow was found to be very small. In this paper, we will test the radial flow effects in much larger systems where the flow effect is presumably much larger, and also the experimental data are available \cite{Kunde1995}. And we will show that a moderate radial flow will have little effect on the cluster formation process in the current model. To attribute the different patterns found in central collisions to radial flow motion, we need to have radial flow energies much larger than the total energy of the collision system. 

For completeness of discussion, we will briefly review the microcanonical lattice gas model \cite{Das2000}, and the incorporation of radial flow in the model (see also \cite{Das2001}). Then we will try to reproduce the experimental data from central and peripheral collisions in the lattice gas model without any flow. We also present the calculations with quite large radial flow. A summary of the essential finding in this work is given in the end.

%
%
%
%

The idea of the lattice gas model is to discretize space and put the nucleons on the lattice \cite{Dasgupta:1995LGM,Dasgupta2001}. Each lattice has at most one nucleon to simulate the short range repulsion effect between nucleons and also partially Pauli principle. The attractive NN interaction is simulated with a bond between a pair of neighboring nucleons, where the bond strengths are take as $\epsilon_{np}=-5.33$ MeV, $\epsilon_{nn}=\epsilon_{pp}=0$. These bond values could roughly reproduce the average binding energy of clusters and the energy density of infinite nuclear matter. The phase space factor is difficult to calculate directly, and is usually sampled with a Metropolis Algorithm. We are using the microcanonical ensemble so the total energy $E_{tot}$ is constant. We start with a given configuration, hence a given $E_{pot}$. Thus the kinetic energy for a given $E_{tot}$ is $E_{kin}=E_{tot}-E_{pot}$. The available phase space of $A=N+Z$ nucleons have this kinetic energy is known analytically: 
\begin{eqnarray}
\Omega(E_{kin}) = \int \delta (E_{kin} - \sum_i \frac{p_i^2}{2 m})\,\prod_i d^3 p_i = \frac{\pi^{3A/2}}{\Gamma(3A/2)} (2m)^{3A/2} E_{kin}^{3A/2 -1}\, .   
 \label{eqn:1}
\end{eqnarray}
We call this value the constant energy integral $\Omega(E_{kin})$. We now attempt a switch in configuration space. The potential energy will change to $E_{pot}'$. To conserve energy, the kinetic energy of the system should be $E_{kin}'=E_{tot}-E_{pot}'$. If $\Omega(E_{kin}')/ \Omega(E_{kin})>1$, the switch is accepted. If it is less than 1, the probability of accepting such a switch is given by the ratio. Since all configurations have equal weights, this satisfies the principle of detailed balance. After many such switching attempts, an event is accepted. We now have to assign momenta to the nucleons so that the total kinetic energy of the $A$ nucleons add up to correct kinetic energy. This can be done by standard Monte-Carlo sampling (for further details, see \cite{Dasgupta2001}).

The inclusion of radial flow is quite straight forward in lattice gas model. Radial flow is a collective motion and does not change the thermal motion. So we may add the flow motion at the end of configuration sampling. Each nucleon acquires additional momentum $\vec{p_f}'(i) = c [ \vec{r}(i) - \vec{r}_{cm} ] $, and the constant $c$ is adjusted to give the required flow energy. With radial flow, the total energy is now $E_{tot}=E_{therm}+E_{pot}+E_{flow}$. From this, it seems that flow has only made the part of the total excitation energy unavailable for thermal motion. But this is not totally true. The flow motion also affect the cluster formation in the current model.

After the sampling of configurations and momentum assignment, we can search for clusters. Two neighboring nucleons are considered to belong to the same cluster when the energy for this pair is lower than zero: 
\begin{equation}
E_{pair}=\frac{p_r^2}{2 \mu} + \epsilon < 0 \,,
 \label{eq:cluster} 
\end{equation}
where $\vec{p}_r$ is the relative momentum and $\epsilon$ is the bond between the pair. 
This clusterization prescription is standard in Lattice Gas Model \cite{Dasgupta:1995LGM,Campi:1997,Chomaz:1998tp}, and was shown to be similar to the Coniglio-Klein prescription \cite{Dasgupta:1995LGM,Dasgupta2001,Campi:1997}. A cluster formed under this condition is, by definition, stable against particle emissions and can be compared with the charge yields from heavy-ion reactions \cite{Dasgupta2001,Chomaz:1998tp}. The particle stable property of this clusterization method was also verified in a molecular dynamical simulation \cite{Dasgupta:1996LGM}.  

In the pair definition, Eq.(\ref{eq:cluster}), the relative motion has components from radial flow, so that in general, pairs are less bound than without radial flow. This effect tends to reduce the cluster size formed on the lattice. On the other hand, increasing thermal motion also has the effect of reducing the binding between neighboring nucleons and subsequently reduces the cluster size.
So the question we have to ask is, qualitatively how big is the radial flow effect relative to the thermal effect.

%
%
%
%

\begin{figure}[tbph!]
\center
\includegraphics[scale=0.4]{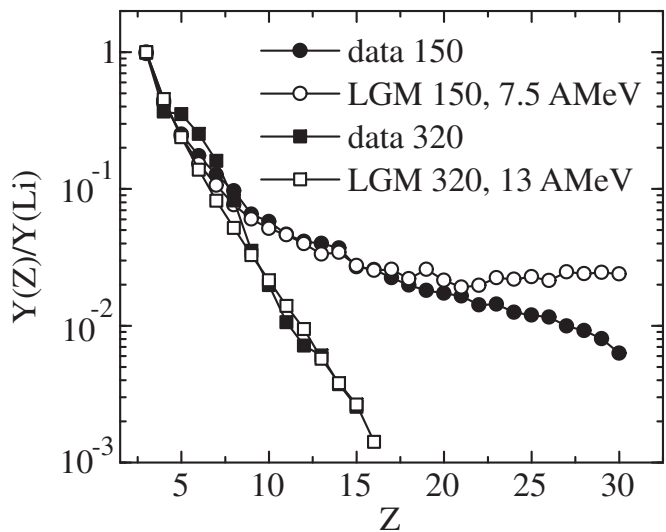}
\caption{
The charge yields from experiment and from lattice gas model for two different systems. The data from experiment (\cite{Kunde1995}) are represented by full symbols, while the results from lattice gas model are shown with open symbols. The two systems used in our simulations are, one with $A=150,\ Z=60,\ N=90$, $E_{therm}=7.5$ AMeV on $8^3$ lattice and one with $A=320,\ Z=130,\ N=190$, $E_{therm}=13$ AMeV on $10^3$ lattice respectively. The charge yields for the $A=150$ system show a typical power law behavior and that for the $A=320$ system show a exponential behavior. All charge yields are normalized to the charge yield of $Li$ element.} 
 \label{fig:1}
\end{figure}   
               
Here we will present the results from microcanonical lattice gas model in two large systems, for which the experimental data were reported before \cite{Kunde1995}. The two systems we used are: $A=150,\ Z=60,\ N=90$ on $8^3$ lattice which corresponds to peripheral $Au+Au$ collisions at $E_{lab}=1$ GeV/nucleon; $A=320,\ Z=130,\ N=190$ on $10^3$ lattice which corresponds to central $Au+Au$ collisions at $E_{lab}=100$ MeV/nucleon. First, we will try to reproduce the experimental data in the lattice gas model with only thermal motion and no radial flow. As shown in Fig.\ref{fig:1}, the full symbols are data for the two system while the open symbols are the corresponding simulation results. Overall, the lattice gas model can reasonably produce the general features of the data. 

For Lattice gas results on $A=150$ system with $E_{therm}=7.5$ AMeV thermal energy, the charge yields show a power law behavior, and closely follow the trend of data except yields for the highest charges. We find the lattice gas model overestimates the charge yields when the cluster size is approaching half of the system size, $Z=30$; this maybe due to edge effect. For the $A=320$ system, the lattice gas results show a nice exponential behavior for charge yields at a higher excitation energy $E_{therm}=13$ AMeV. Even though the energy scale correspondence in the lattice gas model does not correspond exactly to that in a real nuclear system, we find these thermal energies are still in reasonable agreement with expectations. For peripheral reaction as used in the experiment ($A=150$ system), typical excitation energy is estimated to be around $6-8$ MeV (see \cite{Kunde1995} and reference therein). For central collisions, typical excitation energy is expected to be around $15$ MeV.

Following the argument of \cite{Kunde1995}, we have tried to find a reasonable fit to the data on $A=320$ system by fixing the thermal energy to $7.5$ AMeV while changing the radial flow energy. But no matter what reasonable flow energies we used the charge yields are almost not changed within statistical error in the simulations. This is a surprising result. As we will see, random thermal motions are order of magnitude more efficient in breaking the bonds between neighboring nucleons than radial flow motion.

\begin{figure}[tbph!]
\center
\includegraphics[scale=0.4]{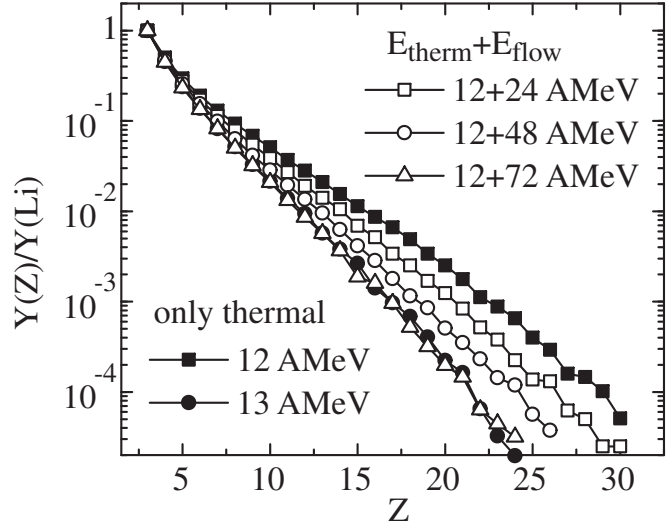}
\caption{
The charge yield for the system $A=320,\ Z=130,\ N=190$ on $10^3$ lattice in the microcanonical lattice gas model. Filled symbols are for simulations with only thermal motion and no radial flow at all, at $E_{therm}=12$ and $13$ AMeV respectively. Open symbols are for simulations with thermal energy fixed at $E_{therm}=12$ AMeV while radial flow energy is varied between $24$, $48$, and $72$ AMeV. As can be seen, the charge yields from simulation with $E_{therm}+E_{flow}=12+72$ AMeV roughly match the charge yields with pure thermal motion $E_{therm}=13$ AMeV. All charge yields are normalized to that of $Li$ element.   
} 
 \label{fig:2}
\end{figure}

The effect on charge yields from radial flow and thermal motion are shown in Fig.\ref{fig:2}. For the charge yields from the lattice gas model, the difference between $12$ and $13$ AMeV pure thermal motion is very large. If we want to produce the charge yields of $E_{therm}=13$ AMeV by introduce a large radial flow on the $E_{therm}=12$ AMeV results, we need about $72$ AMeV radial flow. In other words, the random thermal motion is $72$ times more efficient than radial flow motion at breaking the bonds between neighboring nucleons. 

The reason flow affects clusterization so little is because bond is between nearest neighbors.  For a bond to break we must have large $(\vec p_1-\vec p_2)$.  With flow we add an extra momentum difference of $c(\vec r_1-\vec r_2)= c\vec l$ where $|l|=(6.25)^{1/3}fm$.  This is rather small and does not depend upon how big the dissociating system is.  We will now show that $E_{therm}$ is much more efficient in breaking bonds than $E_{flow}$. We do this by first calculating the average value of $(\vec p_1-\vec p_2)^2$ with pure $E_{therm}$.  We then consider if instead of $E_{therm}$, the same energy was put into $E_{flow}$.  We find that the value of $(\vec p_1-\vec p_2)^2$ falls by about a factor of $50$.

For thermal motion $\langle (\vec p_1-\vec p_2)^2 \rangle=2 \langle \vec p_1^2 \rangle =2\langle \vec p^2 \rangle$. It is reasonable that 
\begin{eqnarray}
 \langle \vec p^2 \rangle=2m\frac{E_{therm}}{A} \, .
 \label{eq:2} 
\end{eqnarray} 
where $A$ is the number of nucleons in the system.  The validity of this equation can be established for our model by turning the discrete problem into a continuous one.  Then 
\begin{eqnarray*}
 E_{therm} &=&\kappa\int_0^P4\pi \frac{p^2}{2m}p^2dp  \, ,
 \nonumber  \\
 A &=& \kappa\int_0^P4\pi p^2dp  \, ,
  \nonumber \\
 \langle p^2 \rangle &=& \frac{\int_0^Pp^4dp}{\int_0^Pp^2dp} \, .
\end{eqnarray*}

We now want to estimate momentum mismatch when $E_{therm}=E_{flow}= \sum_i\frac{c^2r_i^2}{2m}$.  Turn this discrete problem again into a continuous problem.  We assume the dissociating system is uniformly distributed in $3$ times the normal volume (A more realistic density distribution is given in \cite{Samaddar}).  Calling this expanded radius $R$ and the reduced density $\rho$ we find \begin{eqnarray*}
 E_{therm}=E_{flow}=\frac{c^2}{2m}4\pi\rho\frac{R^5}{5}
\end{eqnarray*}
This leads to $c^2={10mE_{therm}}/{3AR^2}$.
Comparing $2\times {2mE_{therm}}/{A}$ with $c^2 l^2$, we find the former is larger by a factor $45$ if the dissociating system has $320$ nucleons. A more elaborate estimate using the geometry of the lattice will give a factor of $50$.

%
%
%
%

From our simulation, we see this factor is $72$ instead of $50$, an increase of $44\%$. In fact, the above estimate only gives a lower limit for this relative efficiency, any deviation from a uniform distribution of nucleons in the simulation will definitely increase this ratio. Further, the additional thermal energy for the pair of neighboring nucleons is actually weighted with a distribution for the pair of neighboring particles, so that resulting average will be higher than our estimation. The overall increase of $44\%$ in the ratio of relative bond-breaking efficiency is most likely the result of a combination of these factors. 

Since the flow values we used here are a lot higher than the total energy deposition in central collisions at $E_{lab}=100$ MeV/nucleon, the lattice gas simulations suggests that there is little effect of radial flow on the clusterization process at intermediate energies. The main effect that caused the charge yields to behave like exponential type rather than power law type is the increase of thermal energy in central collisions.

The inefficiency of radial flow on breaking clusters is also observed by Gulminelli and Chomaz \cite{Gulminelli:2004ch} in a canonical lattice gas model. For other commonly used thermal models, which have the same assumption of sudden dissociation, a simple inclusion of radial flow as we did here will not likely produce a significant difference. Different assumptions about the cluster formation may change the relative efficiency of the radial flow and thermal motion in breaking up clusters, but the large difference between radial flow and thermal motion will hold in any reasonable models. 

The above discussion did not address the issue of the dynamical evolution of the heavy-ion reaction system, especially during the violent stage of the reaction. The clusterization method used in Lattice Gas Model is similar to the $\vec{r}-\vec{p}$ phase space coalescence. If we blindly apply the estimate to normal density, we would find that thermal motion is $22$ times more efficient in breaking bonds than radial flow motion.  Thus we may expect our idea to shed some light on the violent stage of the reaction as well. The dynamical models has also made significant progress to in reproducing cluster formations. In QMD model, algorithms like ECRA and SACA have been formulated to identify asymptotic clusters at an early stage \cite{Dorso:1993,Chernomoretz:2004,Puri:1998te}. In BUU model, stochastic Langevin method was introduced to address the stochastic nature of cluster formations \cite{Colonna:1998}. These methods are calculational intensive. We hope future dynamical simulations will confirm our expectations.

%
%
%
%

To summarize, we have tested the effect of radial flow on cluster formation in microcanonical lattice gas model. We used quite large system where the radial flow effect is expected to be strong and data already existed. We found that reasonable amount of radial flow could not significantly change charge yields in the simulation. For the two systems as discussed in \cite{Kunde1995}, we found the charge yields changed from power law behavior to an exponential behavior purely because of increased thermal energy. Thermal energy is order of magnitude more efficient in reducing the size of large clusters than radial flow. These conclusions should be valid for most thermal models where dynamical evolution of the source is ignored.

%
\section{
    Acknowledgments
}
%
This work is supported in part by the Natural Sciences 
and Engineering Research Council of Canada and by Fonds Nature et Technologies of Quebec.    

%
%
%
%
%
%
%

%
%
%
%
%
\end{document}